\documentclass[acmsmall,screen,nonacm]{acmart}

\AtBeginDocument{%
  }

\usepackage{xspace}
\usepackage{enumitem}
\usepackage[linesnumbered,vlined]{algorithm2e}
\usepackage{listings}
\usepackage{fancyvrb}
\usepackage{pifont}
\usepackage{multirow}
\usepackage{booktabs}
\usepackage[table]{xcolor}

\newcommand{\tool}[0]{\textsc{CapNVI}\xspace}
\newcommand{\pnvi}[0]{PNVI\xspace}
\newcommand{\pnviany}[0]{PNVI-$*$\xspace}
\newcommand{\pnviplain}[0]{PNVI-plain\xspace}
\newcommand{\pnviae}[0]{PNVI-ae\xspace}
\newcommand{\pnviaeudi}[0]{PNVI-ae-udi\xspace}

\newcommand{\myparagraph}[1]{\vspace{0.35em}\noindent\emph{#1.}}
\newcommand{\cmark}[0]{\color{green!50!black}{\ding{51}}}
\newcommand{\xmark}[0]{\color{red!50!black}{\ding{55}}}

\renewcommand{\equationautorefname}{\relax}
\def\equationautorefname#1{}

\definecolor{darkgreen}{RGB}{0,100,0}
\definecolor{darkred}{RGB}{100,0,0}
\definecolor{darkblue}{RGB}{0,0,100}
\definecolor{darkyellow}{RGB}{100,100,0}

\begin{document}

\title{Strict Capabilities with a PNVI Escape Hatch}

\author{Yihe Li}
\orcid{0009-0008-2257-0406}
\affiliation{%
  \institution{National University of Singapore}
  \city{Singapore}
  \country{Singapore}
}
\email{yihe.li@u.nus.edu}

\author{Sai Dhawal Phaye}
\orcid{0009-0005-5960-0574}
\affiliation{%
  \institution{National University of Singapore}
  \city{Singapore}
  \country{Singapore}
}
\email{sdp@u.nus.edu}

\author{Roland H. C. Yap}
\orcid{0000-0002-1188-7474}
\affiliation{%
  \institution{National University of Singapore}
  \city{Singapore}
  \country{Singapore}
}
\email{ryap@comp.nus.edu.sg}

\author{Trevor E. Carlson}
\orcid{0000-0001-8742-134X}
\affiliation{%
  \institution{National University of Singapore}
  \city{Singapore}
  \country{Singapore}
}
\email{tcarlson@nus.edu.sg}

\author{Gregory J. Duck}
\orcid{0000-0002-0837-9671}
\affiliation{%
  \institution{National University of Singapore}
  \city{Singapore}
  \country{Singapore}
}
\email{gregory@comp.nus.edu.sg}

\begin{abstract}
Capability systems enforce memory safety by associating pointers (capabilities) with the allocations from which they were derived---preventing pointer forgery, use-after-free errors, and out-of-bounds memory accesses.
However, existing capability abstractions can also be simultaneously {\em too strict} and {\em not strict enough} for modern \texttt{C} programs.
Existing capabilities can be {\em too strict} for common low-level \texttt{C} programming idioms that manipulate pointers directly through integer casts or raw byte representations.
Likewise, existing capabilities are {\em not strict enough} to enforce all \texttt{C} language pointer rules that are necessary for security---especially regarding {\em type correctness} and {\em sub-object bounds}, whose violation can lead to exploitable behavior.
Modern \texttt{C} provenance models, specifically the proposed {\em Provenance Not Via Integers} (PNVI) extensions, make this tension explicit: ordinary pointer operations must obey strict language-level rules, while low-level manipulation through integers is supported under a deliberately permissive semantics (i.e., a ``PNVI escape hatch'').

In this paper, we argue that these requirements should not be reconciled by strengthening or weakening a single capability abstraction.
Rather, we propose two interoperable capability abstractions.
The first abstraction is \emph{strict capabilities}, which enforce strong language-level pointer invariants, including the provenance, liveness, and object bounds guarantees of existing capability abstractions, as well as {\em type correctness} and {\em sub-object bounds} as stricter extensions.
Capability operations maintain these invariants, preventing invalid pointers from being constructed or dereferenced.
The second abstraction is {\em exposed representations}, which are non-dereferenceable, integer-like values governed by the \pnvi semantics---supporting integer and byte-level manipulation while retaining unforgeability guarantees.
Crucially, exposed representations do not directly encode strict capability authority: conversion back reconstructs provenance under the \pnvi rules and validates type and bounds before memory access is permitted.
This separation allows programs to maximize the use of strict, type-safe capabilities while explicitly crossing into the relaxed domain only where low-level \texttt{C} compatibility is demanded.
We realize our design using {\em Fully Randomized Pointers} (FRP) and dynamic type metadata, and evaluate it on low-level pointer idioms, Juliet tests, and existing \texttt{C} benchmarks.
Our results show that strict type and sub-object safety can coexist with low-level pointer manipulation, without making ordinary capability derivation permissive.
\end{abstract}

\maketitle

\section{Introduction}
\label{sec:intro}

Memory safety remains a fundamental challenge for systems programming languages such as \texttt{C}.
Capability systems~\cite{cheri, yu2023capstone} provide a strong abstraction that helps to address this problem: rather than treating a pointer as merely an integer address, a capability tightly couples the pointer representation with the allocation from which the pointer was derived, and also restricts its authority over memory access.
Modern capability architectures can thereby enforce properties---such as {\em pointer provenance}, {\em object bounds}, and {\em liveness}---preventing important classes of exploitable memory error, including {\em pointer forgery}, {\em out-of-bounds access}, and {\em use-after-free}.

Despite strong provenance/bounds/liveness guarantees, existing capability abstractions nevertheless do not capture all \texttt{C} language-level rules that are necessary to determine pointer validity.
Notably, \texttt{C} imposes strict {\em type rules} on memory access that are independent of bounds and liveness, requiring that the type used by memory access be consistent with the type of the original allocation.
From the perspective of the \texttt{C} abstract machine, type violations are a form of {\em undefined behavior} (UB).
Furthermore, type violations are the basis of \emph{type confusion attack}---a well known class of security vulnerability~\cite{kell16type, haller16typesan, lee15caver, duck18effective}.
For example, the following is potential type confusion:

{\small
\begin{Verbatim}[commandchars=\\\{\}]
    \textcolor{darkgreen}{struct} User   \{ \textcolor{darkgreen}{int} uid; \textcolor{darkgreen}{char} password[\textcolor{darkred}{28}]; \};
    \textcolor{darkgreen}{struct} Packet \{ \textcolor{darkgreen}{int} len; \textcolor{darkgreen}{char} payload[\textcolor{darkred}{28}]; \};
    sendPacket((\textcolor{darkgreen}{struct} Packet *)user, \textcolor{darkyellow}{sizeof}(\textcolor{darkgreen}{struct} Packet));  \textcolor{darkblue}{// password disclosure}
\end{Verbatim}
}

\noindent
Here, the \texttt{password} is confused with a \texttt{payload}, leading to a potential information disclosure attack.
The \texttt{user} pointer possesses valid provenance and is within bounds, while the access itself uses the wrong type (\texttt{User} versus \texttt{Packet}).
Existing capability systems generally do not enforce effective-type correctness, leaving a fundamental gap in their protection against type-confusion vulnerabilities.
Furthermore, traditional capability systems often disable sub-object bound narrowing by default~\cite{zaliva24cheri}, further weakening security.

One natural conclusion might be that {\bf capabilities are not strict enough}.
That is, in addition to enforcing provenance, object-bounds, and liveness, the capability abstraction should also enforce the precise {\em type} and {\em sub-object bounds information} to which a pointer refers.
Pointer operations can therefore preserve these invariants, and reject operations whenever the result cannot be represented as a valid pointer.
Attempts to access a capability using an incompatible type or outside the sub-object bounds will trap, preventing type confusion attacks and fine-grained memory safety.
Under this interpretation, a capability is not merely an authorization to access a range of addresses: it represents a pointer satisfying a collection of more precise language-level safety invariants.
We refer to this interpretation of pointers as \emph{strict capabilities}.

Unfortunately, this approach also exacerbates a different long-standing problem: namely, {\bf capabilities are already too strict}.
Low-level \texttt{C} programs do not always manipulate pointers as pointers.
They routinely expose pointer representations by converting pointers to integers, copying them  through byte arrays, or applying integer and bitwise operations before reconstructing a final pointer.
Such operations appear in systems-code idioms including: {\em tagged pointers}, {\em XOR-linked lists}, {\em serialization}, and other low-level runtime code.
For example, an {\em XOR-linked-list} is a doubly-linked list where the forward- and back-pointers are XOR'ed together---saving a word of storage.
Assuming the \texttt{prev} pointer is known, the list can be traversed in the forward direction as follows:

{\small
\begin{Verbatim}[commandchars=\\\{\}]
                next = (\textcolor{darkgreen}{struct node *})((\textcolor{darkgreen}{uintptr_t})prev ^ curr->npx);
\end{Verbatim}
}

\noindent
A strict capability abstraction treats \texttt{next} as a \emph{forged} capability and rejects it, despite the operation being otherwise safe.
As such, existing capability systems are often incompatible with low-level idioms, requiring source code to be rewritten to remove such idioms~\cite{wang24cheri}.

Existing capability abstractions therefore face an apparent paradox: they are simultaneously \emph{not strict enough} and \emph{too strict} for many \texttt{C} programs.
Specifically, they are not strict enough to enforce safety-critical invariants such as type correctness and sub-object bounds, and yet they are too strict to accommodate intentional representation-level manipulation commonly used by \texttt{C} programs.
Weakening the capability abstraction to admit these idioms undermines the invariants capabilities are expected to enforce, and strengthening the abstraction to capture additional language rules makes the compatibility problem worse.
We argue that these requirements {\em should not be reconciled within a single capability abstraction}.
Instead, the central challenge is how to separate the freedom to manipulate a pointer's representation from the authority to access memory, without allowing the former to undermine the latter.

We argue that the solution to this paradox comes from modern developments in \texttt{C} program semantics.
Specifically, the proposed \emph{Provenance Not Via Integers} (PNVI) extensions~\cite{pnvi} are designed to distinguish ordinary pointer operations (e.g., pointer arithmetic) from those that expose and directly manipulate pointer representations.
Within the strict domain, provenance/bounds/liveness/types constrain what object may be accessed and how.
Within the exposed domain, code may perform integer and byte-level operations without propagating pointer provenance through every intermediate value.
Instead, provenance is {\em reconstructed} on demand whenever a representation is converted back into a pointer, subject to the rules of the PNVI provenance model.
Thus, strengthening capabilities need not require abandoning low-level pointer manipulation:
capabilities can be stricter where \texttt{C} requires strictness, and permissive where \texttt{C} explicitly permits representation-level manipulation.

This paper applies this separation to capability systems in the form of \tool.
Rather than compromising between strong safety and compatibility within a single capability abstraction, \tool separates pointer representations from memory-access authority.
{\em Strict capabilities} enforce provenance, liveness, type correctness, and (sub-)object bounds, and all operations over strict capabilities preserve these invariants.
{\em Exposed representations}, by contrast, are a non-dereferenceable, integer-like representation that may undergo arbitrary integer and byte-level manipulation.
Such representations carry no direct exercisable memory-access authority: conversion back to a strict capability requires reconstruction and validation.
This separation allows strict capabilities to be stricter, and escapes to target specific idioms, illustrated as follows:

\vspace{0.5em}
\begin{center}
{\small
\begin{tabular}{@{}llll@{}}
\toprule
\textbf{Example}
  & \textbf{Traditional}
  & \textbf{C + PNVI}
  & \textbf{CapNVI} \\
\midrule
Type-confused access
  & \textcolor{darkred}{\xmark~Allows}
  & \textcolor{darkgreen}{\cmark~Rejects}
  & \textcolor{darkgreen}{\cmark~Rejects (type error)} \\
XOR-linked list
  & \textcolor{darkred}{\xmark~Rejects}
  & \textcolor{darkgreen}{\cmark~Allows}
  & \textcolor{darkgreen}{\cmark~Allows (exposed)} \\
\bottomrule
\end{tabular}
}
\end{center}
\vspace{0.5em}

\noindent
Unlike traditional capability systems, which are simultaneously too strict and not strict enough, \tool is aligned with the actual \texttt{C} and PNVI semantics.

A central technical challenge remains: {\em how can a capability expose its representation to arbitrary integer and byte-level manipulation without compromising the enforced authority?}
Traditional capability systems protect authority by embedding a representation of authority in the capability representation itself, and rely on special hardware tags to prevent forgery or unauthorized modification.
This approach is fundamentally at odds with the direct representation manipulation permitted by \pnvi.
Our key observation is that the representation need not directly encode the capability's authority.
Instead, an exposed representation need only identify an allocation, and the authority of a reconstructed strict capability can be validated independently against allocation metadata and the \texttt{C} pointer rules.
We realize this separation using {\em Fully Randomized Pointers} (FRP)~\cite{phaye25frp}, whose very high-entropy representations permit provenance reconstruction without propagating metadata through arbitrary integer computations.
EffectiveSan-style dynamic type metadata~\cite{duck18effective} then enables the reconstruction and validation of type and (sub-)object bounds.
Consequently, arbitrary manipulation of exposed representations cannot directly confer memory-access authority, while the randomized encoding provides overwhelming probabilistic resistance to forging representations of otherwise unavailable allocations.
This approach eliminates the need for capability tags or per-pointer authority metadata.
In summary, this paper makes the following contributions:
\begin{itemize}[leftmargin=*]
\item We identify a fundamental tension in applying capability abstractions to \texttt{C}: conventional capabilities do not fully enforce language-level type and sub-object bounds, yet restricting pointer authority more aggressively conflicts with common low-level programming idioms.

\item We propose \tool, a dual abstraction that separates \emph{strict capabilities}, enforcing provenance, liveness, effective type, and sub-object bounds, from non-dereferenceable \emph{exposed representations}, which permit integer- and byte-level pointer manipulation.
Explicit, checked recovery connects the two domains without weakening the monotonicity of ordinary strict capability operations.

\item We show how to realize this separation without
per-pointer authority metadata or capability tags. Our
instantiation combines {\em Fully Randomized Pointers} (FRP), for computationally unforgeable representations, with EffectiveSan-style dynamic type metadata and canonical sub-object bounds---validating recovered authority while preserving strict pointer derivation.

\item We implement a software prototype of \tool and evaluate its safety and compatibility properties.
Tests of low-level pointer idioms and Juliet programs demonstrate both representation-level flexibility and enforcement of strict type, lifetime, and sub-object bounds; and experiments with existing \texttt{C} benchmarks demonstrate the feasibility of adapting substantial programs to the abstraction.
\end{itemize}

\section{Background}

\subsection{Type and Memory Safety in \texttt{C}}\label{sec:safety}

The \verb+C+ programming language~\cite{c23} requires that all memory access obey certain rules in order to be valid.
Each access must be to a {\em live} object (no {\em use-after-free}), the access must be within object bounds (no {\em out-of-bounds} access), and the access must be consistent with the object's {\em effective type} (no {\em type error}, no {\em type confusion}).
Any violation of these rules is expressly {\em undefined behavior} (UB), meaning that the \verb+C+ standard does not prescribe any execution semantics.

Modern \verb+C+ compilers optimize programs under the assumption that UB does not occur.
If (or when) this assumption is broken, the resulting program may behave unpredictably and may leave the program vulnerable to attack.
For example, an {\em out-of-bounds read} may be used to access information stored in adjacent objects---possibly leading to an {\em information disclosure attack} similar to the famous Heartbleed bug~\cite{heartbleed}.
Similarly, a {\em use-after-free} or {\em type confusion} may be used to overwrite a function pointer stored in a different object---possibly leading to a {\em control-flow hijacking} attack.
The programmer is normally expected to ensure that UB does not occur.
However, this is difficult to achieve in practice, as the defined-ness (and safety) of memory access depends on multiple properties, including {\em pointer provenance}, {\em object liveness}, {\em array bounds}, and runtime {\em type information}, meaning that mistakes can be easily made.
We briefly review these properties here.

\subsubsection{Pointer Provenance}

The concept of \emph{pointer provenance}~\cite{memarian19provenance} captures the idea that pointers are uniquely associated with the {\em allocation} (a.k.a. ``{\em storage instance}'') from which the pointer was derived.
Although \verb+C+ programmers typically think of pointers as raw integer addresses, the \texttt{C} standard implicitly treats pointers as richer values: conceptually as abstract {\em pairs} of values $\langle\pi,\mathit{addr}\rangle$, comprising a numeric address ($\mathit{addr}$) and a {\em provenance tag} ($\pi$) that identifies a unique storage instance.
Standard pointer operations (pointer arithmetic, address-of, etc.) preserve provenance tag $\pi$, meaning that pointers are forever associated with a single storage instance.
The provenance tag $\pi$ is usually implicit, and is important for compiler optimization and alias analysis (two pointers with different provenances cannot legally refer to the same object).
For memory access to be valid, the dereferenced pointer must have a valid provenance to a live (i.e., not \texttt{free}-ed) object.

We highlight that there are several competing models for provenance.
The existing \verb+C+ standard implies a ``{\em strict provenance model}'', meaning that only pointers derived from well-defined operations (e.g., pointer arithmetic) will have a valid provenance.
Pointers derived by other means (such as byte-wise pointer manipulation or integer-pointer casts) are generally UB to dereference.
However, in recent years, there has been a growing recognition that the strict provenance model is too strict, and excludes many real-world programming idioms that are generally compiled safely.

\subsubsection{Array and (Sub-)Object Bounds}

Under the \verb+C+ standard, pointer arithmetic is only well-defined within the bounds of an array object, and a special case of {\em one-past-the-end} of the last element~\cite{c23}~\S6.5.6.
Dereferencing a pointer is only well-defined if the pointer points to an actual element of the array itself.
Accessing outside the array bounds (including the one-past-the-end pointer) is UB.
The \verb+C+ standard treats a single non-array object as an array of length one for the purpose of these rules, so pointer arithmetic and validity checks apply in the same manner.
When dealing with sub-objects (such as members of a \verb+struct+ or elements of a nested array), the bounds apply to the containing sub-object treated as an array of its own size, and not to the enclosing object as a whole.
Again, violating any of these constraints is UB.

\subsubsection{Dynamic (Effective) Types}

Each object in \verb+C+ has a \emph{dynamic type}, a.k.a., the object's \emph{effective type}~\cite{c23}.
The effective type determines what kinds of pointer operations are valid.
In particular, the effective type constrains pointer dereferences: if an lvalue expression of type \texttt{T} designates an object whose effective type is incompatible with \texttt{T}, then the behavior is undefined.
Similarly, pointer arithmetic is also type-directed: advancing a (\texttt{T *}) pointer by $n$ steps is only meaningful when the underlying storage consists of contiguous \texttt{T} objects.
Pointer casts do not change the effective type of the underlying object.
For example, converting a (\texttt{T *}) to a (\texttt{U *}) is permissible, but using the resulting pointer to access the object as a \texttt{U} may be undefined unless \texttt{U} is compatible with \texttt{T}.

The primary exemption is for (\texttt{char}) and (\texttt{unsigned char}), which may alias any object representation: a (\texttt{char *}) may safely be used to examine or copy the underlying bytes of any object.
Beyond type compatibility, the standard also imposes alignment requirements: a pointer must be suitably aligned for the type to which it points.
For example, given the program fragment:

{\small
\definecolor{darkgreen}{RGB}{0,100,0}
\definecolor{darkred}{RGB}{100,0,0}
\begin{Verbatim}[commandchars=\\\{\}]
                \textcolor{darkgreen}{int} a[\textcolor{darkred}{10}]; \textcolor{darkgreen}{int} *p = (\textcolor{darkgreen}{int} *)((\textcolor{darkgreen}{char} *)a + \textcolor{darkred}{1}); *p = \textcolor{darkred}{3};
\end{Verbatim}
}

\noindent
the assignment through \texttt{p} is UB since the alignment constraints of \texttt{int} are not satisfied.

\subsubsection{Other Cases}
Beyond provenance, bounds, and effective type, several additional rules can affect the definedness of pointer dereference.
For example, special rules exist for \texttt{const}-, \texttt{restrict}-, and \texttt{volatile}- qualified pointers, and {\em unsynchronized} concurrent modification or access may lead to a data race, which is also UB.
\emph{Uninitialized reads} are also generally UB, or else the read returns an \emph{indeterminate value} in the case of (\texttt{char}) or (\texttt{unsigned char}).

\subsubsection{Enforcing Pointer Safety.}
The pointer properties described above are largely implicit in the \texttt{C} abstract machine, with violations treated as UB.
That said, several systems attempt to make some of these properties explicit.
Software sanitizers, such as AddressSanitizer~\cite{asan}, SoftBound~\cite{nagarakatte09softbound}, and LowFat~\cite{duck16heap}, track object bounds or allocation validity to detect spatial and/or temporal memory errors.
Hardware-supported capability systems, such as CHERI~\cite{cheri} and Capstone~\cite{yu2023capstone}, go further by representing pointers as {\em unforgeable} capabilities with associated provenance, bounds, and permissions.
However, these mechanisms generally do not enforce \texttt{C}'s effective-type rules, meaning that
a memory access may be permitted by a capability while
being UB due to type
confusion.

Other systems provide complementary mechanisms.
EffectiveSan~\cite{duck18effective} uses dynamic type metadata to detect type errors and reconstruct sub-object bounds from an object's type and layout.
{\em Fully Randomized Pointers} (FRP)~\cite{phaye25frp} instead use randomized pointer representations to associate pointers with allocations, providing {\em probabilistic unforgeability} and spatial/temporal checking.
These mechanisms illustrate that provenance, bounds, liveness, and type correctness can be enforced using different representations and metadata schemes.
However, enforcing these properties does not by itself resolve the separate problem of compatibility with low-level pointer representation manipulation, which motivates the provenance models discussed next.

\subsection{Modern Provenance Models}\label{sec:prov-models}

Since there exists a tension between the \emph{strict provenance model} (implied by the \verb+C+ standard) and real-world systems code, there has been a push to develop more realistic provenance models.

\subsubsection{Provenance Not Via Integers}

The {\em Provenance Not Via Integers} (\pnvi) models have gained significant traction. 
These differ from the strict provenance model in that they allow pointer provenance to be {\em reconstructed} based on the bit pattern of a pointer value.
For example, suppose that a pointer \verb+p+ is created using a cast from an integer \verb+i+, e.g., (\verb+p = (int *)i+), and pointer \verb+p+'s bit pattern matches a {\em valid address} for currently live storage instance $\mathit{obj}$.
Then the \pnviany model may allow for \verb+p+'s provenance to be {\em reconstructed} to that of $\mathit{obj}$ provided certain conditions are met.
As such, the \pnvi models can assign well-defined behavior even to low-level pointer operations, and more closely match the intended semantics of low-level systems code.

There is more than one \pnvi model, including  \pnviplain~\cite{pnvi}, \pnviae (\emph{address exposed})~\cite{pnvi}, \pnviaeudi (\emph{address exposed with user disambiguation})~\cite{pnvi, pnvi-ae-udi}.
These models differ mainly on which provenance reconstructions are considered valid.
\pnviplain is the most liberal, allowing provenance to be reconstructed for a live storage instance whose address happens to match the pointer's bit pattern.
\pnviae further restricts provenance reconstruction based on the notion of \emph{exposure}.
Intuitively, a storage instance becomes {\em exposed} when the \verb+C+ abstract machine can no longer guarantee that all uses of its addresses are only derived from sanctioned operations, such as ordinary pointer arithmetic.
Once exposed, integer-to-pointer casts or raw byte manipulation, whose bit values that fall within that allocation, may reconstruct its provenance and potentially reintroduce aliasing relationships.
Several operations can expose a storage instance:
\begin{enumerate}[leftmargin=*]
  \item \emph{Pointer-to-integer casts}: converting a pointer to an integer (e.g., \verb+uintptr_t+ or \verb+intptr_t+) exposes the corresponding allocation.
  \item \emph{Representation-byte access}: reading the raw bytes of a pointer (e.g.\ via a cast to (\verb+unsigned char *+) or a \texttt{union}) exposes the pointed-to allocation.
  \item \emph{I/O output}: printing a pointer with \verb+%p+, or writing a raw pointer value to a file, etc., exposes the corresponding allocation.
\end{enumerate}
Exposure is defined per storage instance: once any pointer into an allocation is exposed, the entire allocation is considered exposed for the rest of the execution until the object is freed.
\pnviaeudi further refines the model by handling some ambiguous cases during reconstruction.
This paper adopts \pnvi-style reconstruction, but does not explicitly track the historical exposure state of \pnviae, and ambiguity will be handled explicitly rather than using \pnviaeudi's user disambiguation.
Instead, exposure and ambiguity will be handled via other means.

\lstdefinestyle{CStyle}{
  language=C,
  basicstyle=\ttfamily\tiny,
  keywordstyle=\color{blue},
  morekeywords={ptrdiff_t,intptr_t},
  morekeywords=[2]{malloc,free},
  keywordstyle=[2]\color{teal},
  commentstyle=\color{gray}\itshape,
  stringstyle=\color{red},
  showstringspaces=false,
  numbers=none,
  escapeinside={(*@}{@*)},
  tabsize=2,
  xleftmargin=1em,
  frame=single,
  framesep=2pt,
  backgroundcolor=\color{gray!8},
  literate={0}{{{\color{red}0}}}{1}
           {1}{{{\color{red}1}}}{1}
           {2}{{{\color{red}2}}}{1}
           {3}{{{\color{red}3}}}{1}
           {4}{{{\color{red}4}}}{1}
           {5}{{{\color{red}5}}}{1}
           {6}{{{\color{red}6}}}{1}
           {7}{{{\color{red}7}}}{1}
           {8}{{{\color{red}8}}}{1}
           {9}{{{\color{red}9}}}{1}
}

\lstdefinestyle{CStyleNumbers}{
  style=CStyle,
  numbers=left,
  numberstyle=\tiny\color{gray},
  stepnumber=1,
  numbersep=6pt
}

\newcommand{\hlcode}[1]{\begingroup
  \setlength{\fboxsep}{1pt}
  \colorbox{red!15}{#1}%
\endgroup}

\subsection{Motivating Example: Sub-object Bounds}
\label{sec:example}
Traditional capability systems only support a variant of the {\em strict provenance semantics} implied by the \texttt{C} standard, and not any \pnvi semantics.
As such, this can create incompatibilities with common low-level \texttt{C} idioms, and such incompatibilities {\em are often resolved by disabling strict protections}.
For example, consider the $\texttt{container\_of}(\mathit{ptr},T,\mathit{field})$ idiom that is very common in kernel-level programming \cite{linux-containerof}.
This operation recovers a pointer to an enclosing structure (of type $T$) from a pointer $\mathit{ptr}$
to one of $T$'s $\mathit{field}$ member:

\vspace{0.6em}
{ \small
\begin{tabular}{c|c}
\begin{minipage}{0.64\textwidth}
\begin{Verbatim}[commandchars=\\\{\}]
\textcolor{darkblue}{#define container_of(ptr, T, m) \textbackslash}
\textcolor{darkblue}{  ((T *)((uintptr_t)(ptr) - (uintptr_t)offsetof(T, m)))}
\textcolor{darkgreen}{struct} list_head *p = &obj->link;
\textcolor{darkgreen}{struct} object *q = \textcolor{darkblue}{container_of}(p, \textcolor{darkgreen}{struct} object, link);
\end{Verbatim}
\end{minipage}
&
\begin{minipage}{0.25\textwidth}
\begin{Verbatim}[commandchars=\\\{\}]
 \textcolor{darkgreen}{struct} object \{
   \textcolor{darkgreen}{long} value;
   \textcolor{darkgreen}{struct} list_head link;
 \};
\end{Verbatim}
\end{minipage}
\end{tabular}
}
\vspace{0.6em}

\noindent
Here, \texttt{container\_of} is defined using a pointer-integer round trip and integer arithmetic.\footnote{
The other common definition of \texttt{container\_of} uses a byte pointer (\texttt{char *}) round-trip and pointer arithmetic.
However, this alternative is considered to rely on {\em ambiguous behavior} under the \texttt{C} standard (see~\cite{wg14-n2222} \S 2.5.4).
}
Assuming that \texttt{p} is a capability bound to specifically the \texttt{link} sub-object, then a sub-object bounds error, such as \texttt{((long *)p)[-1]}, will be prevented from accessing the \texttt{value} member.
However, conventional capability derivation is {\em monotonic}: a capability cannot acquire authority beyond its existing bounds.
Consequently, a capability bounded to \texttt{link} cannot ordinarily be transformed into one covering the enclosing \texttt{object}.
Thus, in order to support \texttt{container\_of}, the entire container-wide bounds may need to be retained on member capabilities.
The CHERI \texttt{C}/\texttt{C++} Programming Guide~\cite{cheri2020guide} states: 
\begin{quote}
{\em
Subobject bounds are not enabled by default as they may require additional source code changes for compatibility.}
\end{quote}
This means that {\em sub-object bounds protection is not enforced by default}, regardless of whether the operation is legitimate or not.
In other words, \texttt{((long *)p)[-1]} is {\bf allowed} under this default, despite being a potentially exploitable sub-object bounds error.

\tool separates these concerns.
The member pointer remains a strict, sub-object-bounded, and type-safe capability during ordinary operations.
When \texttt{container\_of} is required, its representation can be exposed and manipulated
under \texttt{C}'s \pnvi semantics.
Reconstruction then recovers the enclosing object's provenance and validates the type and bounds before producing a new strict capability, meaning that compatibility need {\em not} require the universal weakening of ordinary dereferenceable capabilities.
Rather, the relaxation of sub-object bounds is permitted {\em only} when the operations are compliant with the existing \texttt{C} standard and \pnvi provenance reconstruction semantics.
Under this approach, \texttt{((long *)p)[-1]} will be {\bf disallowed} since it is a straightforward violation of \texttt{C}'s semantics, while \texttt{container\_of} is {\bf allowed} under \pnvi provenance reconstruction.

\subsection{Problem Statement}

Existing capability abstractions face two conflicting requirements: strictness versus compatibility.
We argue that these requirements should be separated into two interoperable domains: a \emph{strict} domain that enforces strong pointer invariants, and an \emph{exposed} domain that permits arbitrary representation-level manipulation without directly authorizing memory access.
The central challenge is to make this separation enforceable:
{\em how can a capability system expose its representation to arbitrary manipulation without allowing forgery or illegitimate amplification of the capability's authority?}
In particular, conversion from an exposed representation must recover provenance according to \pnvi reconstruction semantics, while also reconstructing and validating the type, bounds, and liveness invariants of the strict capability domain.
Crucially, this must be achieved without relying on
unforgeable capability tags or other protected
per-pointer metadata, since exposed representations
must remain freely manipulable as ordinary data.

We address this problem by first formalizing the semantics of strict capabilities and exposed representations (\autoref{sec:design}).
We then instantiate the semantics in the form of \tool (\autoref{sec:instantiation}).

\section{Dual Capability-Representation Design} \label{sec:design}

Our goal is to design a capability system that strongly enforces spatial, temporal, and type safety guarantees while also being flexible enough to support a modern \pnvi provenance model for compatibility.
Existing capability machines enforce a strict provenance discipline: every valid pointer must be derived from an existing capability through a small set of {\em sanctioned} operations (pointer-arithmetic, address-of, etc.).
While sound, this approach rejects many idiomatic \verb+C+ patterns.
Our design bridges this gap by supporting {\em two interoperable capability representations} that together realize both the strict and relaxed subsets of \verb+C+.
Our core idea is to represent pointers as a {\em strict} capability (for precise metadata tracking and dereference) or as an {\em exposed representation} (for relaxed raw-byte and integer-like manipulation) depending on the context.

Importantly, the exposed domain is optional: programs may operate entirely using strict capabilities, without invoking \pnvi-style exposure or reconstruction.
This is analogous with the distinction between safe and unsafe Rust, where programs can remain within a restricted, safety-preserving subset, but may explicitly escape when necessary to implement more low-level idioms.
Similarly, \tool makes strict capabilities the default programming abstraction, with exposure and recovery providing an explicit escape hatch for low-level \texttt{C} idioms.

\subsection{Strict Capabilities}\label{sec:fat}

\begin{figure}[t]
\centering
\begin{minipage}{0.96\linewidth}
\footnotesize
\[
\frac{
  \pi \text{ fresh}
  \quad
  \mathit{a} = \mathsf{base}(\pi)
  \quad
  n = k\cdot\texttt{sizeof(}T{)}
}{
  \texttt{(}T~\texttt{*)malloc(}n\texttt{)}
  =
  \langle \pi,T,T,\mathit{a}..\mathit{a}{+}n,\mathit{a} \rangle \qquad \Sigma[\pi] \leftarrow \mathsf{live}
}
\tag{\textsc{Alloc}}
\label{rule:alloc}
\]
\vspace{0.6em}
\[
\frac{
  c=\langle\pi,S,T,\mathit{lb}..\mathit{ub},\mathit{a}\rangle
  \quad
  \mathit{a}=\mathsf{base}(\pi)
  \quad
  \Sigma[\pi]=\mathsf{live}
}{
  \texttt{free(}c\texttt{)} \qquad
  \Sigma[\pi] \leftarrow \mathsf{dead}
}
\tag{\textsc{Free}}
\label{rule:free}
\]
\vspace{0.6em}
\[
\frac{
  c = \langle \pi,S,T,\mathit{lb}..\mathit{ub},\mathit{a}\rangle
  \quad
  \mathit{a}' = \mathit{a} + k\cdot\texttt{sizeof(}S\texttt{)}
  \quad
  \mathit{lb} \le \mathit{a}' \le \mathit{ub}
}{
  c~\texttt{+}~k =
  \langle \pi,S,T,\mathit{lb}..\mathit{ub},\mathit{a}'\rangle
}
\tag{\textsc{Arith}}
\label{rule:arith}
\]
\vspace{0.6em}
\[
\frac{
  c = \langle \pi,S,T,\mathit{lb}..\mathit{ub},\mathit{a}\rangle
  \quad
  U = \texttt{typeof(}S\texttt{.f)}
  \quad
  \mathit{a}' = \mathit{a} + \texttt{offsetof(}S\texttt{,} \texttt{f)}
  \quad
 \mathit{a}'..\mathit{a}'{+}\texttt{sizeof(}U\texttt{)} \subseteq \mathit{lb}..\mathit{ub} 
}{
  \texttt{\&}c\texttt{->f} =
  \langle \pi,U,T,\mathit{a}'..\mathit{a}'{+}\texttt{sizeof(}U\texttt{)},\mathit{a}'\rangle
}
\tag{\textsc{Member}}
\label{rule:member}
\]
\vspace{0.2em}

$\qquad$
\begin{minipage}{0.37\textwidth}
\[
\frac{
  c = \langle \pi,S,T,\mathit{lb}..\mathit{ub},\mathit{a}\rangle
}{
  \texttt{(}U~\texttt{*}\texttt{)}c =
  \langle \pi,U,T,\mathit{lb}..\mathit{ub},\mathit{a}\rangle
}
\tag{\textsc{Cast}}
\label{rule:cast}
\]
\end{minipage}
\hfill
\begin{minipage}{0.49\textwidth}
\[
\frac{
  c = \langle \pi,S,T,\mathit{lb}..\mathit{ub},\mathit{a}\rangle
  \quad
  \mathsf{isValidDeref}(\Sigma, c)
}{
  v = \mathsf{read}(c,S) \quad \text{or} \quad \mathsf{write}(c, S, v)
}
\tag{\textsc{Deref}}
\label{rule:deref}
\]
\end{minipage}

\vspace{1.2em}
\[
\frac{
  c_1 = \langle \pi_1,S_1,T_1,lb_1..ub_1,a_1\rangle
  \quad
  c_2 = \langle \pi_2,S_2,T_2,lb_2..ub_2,a_2\rangle
  \quad
  \mathit{op} \in \{\texttt{==}, \texttt{!=}\}
  \quad
  \mathsf{isValidEq}(\Sigma, c_1, c_2)
}{
  (c_1~\mathit{op}~c_2)
  \;=\;
  (a_1~\mathit{op}~a_2)
}
\tag{\textsc{Eq}}
\label{rule:eq}
\]
\vspace{0.2em}
\[
\frac{
  c_1 = \langle \pi,S_1,T_1,lb_1..ub_1,a_1\rangle
  \quad
  c_2 = \langle \pi,S_2,T_2,lb_2..ub_2,a_2\rangle
  \quad
  \mathit{op} \in \{\texttt{<}, \texttt{<=}, \texttt{>}, \texttt{>=}\}
  \quad
  \mathsf{isValidRel}(\Sigma, c_1, c_2)
}{
  (c_1~\mathit{op}~c_2)
  \;=\;
  (a_1~\mathit{op}~a_2)
}
\tag{\textsc{Rel}}
\label{rule:rel}
\]
\end{minipage}
\caption{{\em Sanctioned operations on strict capabilities}.
Ordinary strict operations preserve or narrow capability authority (they never widen sub-object bounds).
All operations are partial.
The $\mathsf{isValid*}$ helpers check the operation against the \texttt{C} standard rules. 
If the premises of a rule are not satisfied, the operation traps.
\textsc{Arith} permits a one-past pointer ($a'=ub$), but such a pointer can never
satisfy $\mathsf{isValidDeref}$.
}
\label{fig:strict-semantics}
\end{figure}

A {\em strict capability} is an abstract pointer-like object comprising metadata and an underlying address:

\vspace{-1em}
{\small
\begin{align*}
    \text{strict capability} ~:=~ \langle \pi, S, T, \mathit{lb}..\mathit{ub}, \mathit{addr} \rangle
    \tag{\textsc{Strict-Capability}} \label{eq:fat}
\end{align*}
}

\noindent
Here $\pi$ is the {\em provenance} (unique allocation/object identifier), $S$ is the capability's \emph{static type}, $T$ is the object's \emph{effective type}, $\mathit{lb}..\mathit{ub}$ is the {\em (sub-)object bounds} represented as an address range (in bytes), and $\mathit{addr}$ is the actual underlying address.
Strict capabilities support all pointer-like operations.

\myparagraph{Sanctioned operations}
The {\em sanctioned operations} on strict capabilities are defined in \autoref{fig:strict-semantics}.  These include (de)allocation, pointer arithmetic, casts, comparison, address-of-member (get-element-pointer), and dereference (load/store).
Allocation (\autoref{rule:alloc}) creates a fresh, never-before-used provenance $\pi$ which is initially marked $\mathit{live}$.
Subsequent sanctioned operations preserve $\pi$ and the effective type $T$, while deallocation (\autoref{rule:free}) marks $\pi$ as $\mathit{dead}$.
Pointer casts (\autoref{rule:cast}) change only
the static type $S$ and do not change $T$ or bounds.
Pointer arithmetic (\autoref{rule:arith}) preserves the current bounds, while address-of-member (\autoref{rule:member}) may narrow them.
Neither operation may construct an out-of-bounds capability.
Finally, (\ref{rule:eq}) and (\ref{rule:rel}) implement pointer comparison.

This strict interpretation is deliberate: sanctioned operations never widen the authority of an existing capability.\footnote{The \texttt{C} standard allows some limited widening operations via ordinary casts, such as casting a pointer-to-the-first-member to the pointer to the containing object.
However, strict capabilities deliberately implement monotonicity as a policy.}
Our priority is comprehensive enforcement of language-level pointer invariants, including checks on pointer-producing operations rather than only memory accesses, even where this imposes a more restrictive programming discipline.
Programs requiring less restrictive pointer manipulation can instead use \emph{exposed representations}, introduced below.
Finally, dereference (\autoref{rule:deref}) requires the \texttt{C} typing, alignment, liveness, and bounds requirements for the memory access to be satisfied, abstracted by the $\mathsf{isValidDeref}$ predicate.
We discussed these requirements in \autoref{sec:safety}.
Strict capabilities are also assumed to be
{\em unforgeable}: they can only be constructed through sanctioned strict operations or through the checked reconstruction operation defined for exposed representations below.

\subsection{Exposed Representations}
\label{sec:exposed}

Strict capabilities deliberately prohibit operations that can increase authority.
However, this restriction is incompatible with low-level \texttt{C} idioms that manipulate pointer representations directly.
Rather than dropping strict protection in favor of compatibility, as shown in \autoref{sec:example}, our approach is to allow \pnvi-compatible operations through {\em exposure} and {\em reconstruction}.

An \emph{exposed representation} is the ordinary bit representation of a strict capability.
Let:

\vspace{-1em}
{\small
\begin{align}
    r = \mathsf{repr}(\pi,a)
    \label{eq:repr}
    \tag{\sc Representation}
\end{align}
}

\noindent
denote the {\em representation} of a pointer with provenance $\pi$ and address $a$.
Unlike a strict capability, an exposed representation is assumed to carry no directly usable type or bounds authority, and also cannot be dereferenced.
Instead, $r$ is an ordinary integer or sequence of bytes and may be manipulated using arbitrary representation-level operations.
For example, {\em addition}, {\em masking}, XOR, {\em byte-wise} manipulation, etc., may transform $r$ into any other arbitrary representation $r'$.

{\em Exposure} is modeled by the following rule:

\vspace{-1em}
{\small
\begin{align}
\frac{
    c=\langle\pi,S,T,\mathit{lb}..\mathit{ub},a\rangle
    \qquad
    r=\mathsf{repr}(\pi,a)
}{
    \mathsf{Expose}(c)=r
}
\tag{\textsc{Expose}}
\label{rule:expose}
\end{align}
}

\noindent
Here, the (\autoref{rule:expose}) rule abstracts the exposure events of the underlying \pnvi semantics.
For example, pointer-to-integer conversion exposes the corresponding provenance, as can observation of a pointer representation through a non-pointer type.
Merely manipulating an exposed representation does not confer memory-access authority.

To use a representation as a pointer again, it must be {\em reconstructed} as a
strict capability.
Let $\mathcal{T}[\pi]$ give the effective type associated with provenance $\pi$, and let
$\mathcal{B}[S,T,\mathit{offset}]=\mathit{lb}..\mathit{ub}$ denote a sub-object interpretation for the given static type $S$, effective type $T$, and $\mathit{offset}$, that is admissible under the \texttt{C} standard~\cite{c23}.
Then {\em reconstruction} is defined by the following rule:

\vspace{-1em}
{\small
\begin{align}
\frac{
    r=\mathsf{repr}(\pi,\mathit{offset})
    \qquad
    \mathsf{live}(\pi)
    \qquad
    \mathcal{T}[\pi]=T
    \qquad
    \mathcal{B}[S,T,\mathit{offset}]
        =\mathit{lb}..\mathit{ub}
}{
    \mathsf{Recover}(S, r)
    =
    \langle\pi,S,T,\mathsf{base}(\pi){+}\mathit{lb}~..~\mathsf{base}(\pi){+}\mathit{ub},a\rangle
}
\tag{\textsc{Recover}}
\label{rule:recover}
\end{align}
}

\noindent
If no such interpretation exists, reconstruction fails.
Note that the relation $\mathcal{B}$ need not select a unique interpretation: \texttt{C} permits cases in which the same address admits multiple sub-object interpretations.
Common examples include: unions, one-past-the-end pointers, and byte-pointers (\texttt{char *}) pointers into non-byte types. 
The abstract semantics may therefore admit multiple reconstructions.

Importantly, representation manipulation cannot {\em arbitrarily modify the authority fields of a strict capability}.
A computation on $r$ may change the representation so that recovery resolves to a different address or exposed provenance, but the resulting effective type and bounds are reconstructed from $\mathcal{T}$ and $\mathcal{B}$ rather than supplied by the representation.
Hence there is no operation that transforms one strict capability into another with arbitrarily chosen provenance, effective type, bounds, or address.
The only path from an exposed representation back to memory access is through (\autoref{rule:recover}) and the strict capability checks.
This distinction separates \tool from traditional capability systems in which authority metadata is itself part of a potentially manipulable pointer representation.
This usually necessitates additional safety mechanisms (e.g., hardware tags) to prevent direct modification of capabilities.
In contrast, \tool deliberately permits arbitrary computation over exposed pointer bits, but {\em not} arbitrary computation over capability authority.
For example, exposure may deliberately break the monotonicity of strict capability derivation: a transformed representation can reconstruct a capability for a different sub-object.
Such non-monotonic reconstruction is confined to the explicit exposure/recovery boundary under the \texttt{C}/\pnvi semantics, and operations entirely within the strict domain remain monotonic.

\myparagraph{Example}
Consider the call to $\texttt{container\_of}(\texttt{p}, \texttt{struct object}, \texttt{link})$ from \autoref{sec:example}:
\begin{enumerate}[leftmargin=*]
\item First, converting the strict capability \texttt{p} to an integer invokes (\autoref{rule:expose}), yielding {\em representation} $r$.
\item Ordinary integer arithmetic then computes
$r' = r{-}\texttt{offsetof}(\texttt{struct object},\texttt{link})$.
This is an ordinary integer operation and carries no memory-access authority.
\item Finally, converting $r'$ to a (\texttt{struct object *}) invokes (\autoref{rule:recover}), which {\em reconstructs} a new strict capability, including PNVI provenance, effective type, and sub-object bounds.
\end{enumerate}
Crucially, the arithmetic on $r$ cannot directly choose these authority fields: recovery succeeds only if $r'$ identifies a live, exposed object for which $\mathcal{T}$ and $\mathcal{B}$ admit the requested interpretation.
Thus, when \texttt{p} is a genuine pointer to a \texttt{link} member inside a \texttt{struct object}, the usual \texttt{container\_of} computation can recover a strict capability to the enclosing object.
Otherwise the arithmetic itself provides no authority to manufacture the requested capability.

\section{Instantiating a Dual Capability-Representation System}
\label{sec:instantiation}

\autoref{sec:design} defines strict capabilities abstractly as tuples $\langle \pi,S,T,\mathit{lb}..\mathit{ub},\mathit{addr}\rangle$, together with exposed representations $\mathsf{repr}(\pi,\mathit{offset})$.
A direct implementation could explicitly encode this metadata in each strict capability, such as using a fat-pointer representation for example.
However, since \tool also permits the pointer representation to be \emph{exposed} and manipulated, explicitly encoding authority-sensitive metadata would potentially allow that authority to be modified directly, in violation of the abstract semantics.
To address this issue, our main insight is that \emph{strict-capability metadata can be represented implicitly rather than explicitly}.
Specifically, given a pointer representation and a static type $S$, then the effective type $T$, liveness, and sub-object bounds $\mathit{lb}..\mathit{ub}$ information can be {\em reconstructed} on demand.
Thus, strict and exposed pointers can share the same compact physical representation: i.e., the distinction is semantic rather than representational.
Moreover, authority-sensitive metadata remains protected from direct representation manipulation, since such metadata is never explicitly encoded in
the pointer in the first place.

Our approach realizes this design as a concrete refinement of the abstract semantics.
We use hatted symbols for mechanisms that refine their abstract counterparts.
For example, the abstract representation
$\mathsf{repr}(\pi,\mathit{offset})$ is realized by
$\widehat{\mathsf{repr}}(\pi,\mathit{offset},e)$, where $e$ distinguishes otherwise ambiguous one-past-end pointers.
The instantiation combines two mechanisms:
a deterministic sub-object bounds function $\widehat{\mathcal{B}}$, which reconstructs canonical authority from allocation metadata and static type;
and computationally unforgeable representations, which allow provenance to be recovered without explicit per-pointer metadata.
The latter can be realized using \emph{Fully Randomized Pointers} (FRP)~\cite{phaye25frp}.
Strict operations enforce the rules from \autoref{fig:strict-semantics}, potentially rejecting operations whose authority cannot be reconstructed without additional per-pointer state.
Conceptually, the correspondence is:

\vspace{-0.7em}
{\small
\[
\begin{array}{c|c}
\text{Abstract semantics} & \text{Instantiation} \\ \hline
\mathsf{repr}(\pi,\mathit{offset})
  & \widehat{\mathsf{repr}}(\pi,\mathit{offset},e)
    \quad\text{(allocation-resolving handle)} \\
\mathsf{base}(\pi),\ \mathsf{size}(\pi),\ \Sigma[\pi]
  & \text{per-allocation metadata} \\
\mathcal{T}[\pi]
  & \text{recoverable effective-type metadata} \\
\mathcal{B}[S,T,\mathit{offset}]
  & \text{canonical }\widehat{\mathcal{B}}[S,T,\mathit{offset},e] \\
S
  & \text{static program information}
\end{array}
\]
}

\noindent
As noted above, the concrete mechanisms are \emph{refinements} rather than literal representations of every abstract field.
A refinement may preserve additional representation information needed to realize the abstract semantics, deterministically resolve an abstractly ambiguous interpretation, or conservatively reject an operation that cannot be implemented without additional per-pointer state.
Such restrictions remove executions rather than grant additional strict capability authority.
Other aspects of our instantiation rely on strong probabilistic guarantees to reduce the metadata burden while retaining compatibility

\subsection{Reconstructible Capability Invariants}
\label{sec:reconstructible}

The central invariant of our instantiation is that a strict capability's authority can be reconstructed from its representation, static type, and per-allocation metadata, without retaining per-pointer bounds or type metadata.
The representation $\widehat{\mathsf{repr}}(\pi,k,e)$ identifies provenance $\pi$, allocation offset $k$, and an endpoint discriminator $e$.
The metadata $\mathcal{T}[\pi]$ and $\Sigma[\pi]$ provide the effective type and liveness.
It remains to reconstruct sub-object bounds.

\myparagraph{Canonical sub-object bounds}
The abstract sub-object relation $\mathcal{B}[S,T,k]$ from (\autoref{rule:recover}) may admit multiple bounds interpretations for the same static type $S$, effective type $T$, and offset $k$.
Since the concrete representation does not retain which interpretation was chosen, we define a {\em deterministic bounds function} $\widehat{\mathcal{B}}$ by conservatively intersecting all admissible bounds.

Let $\mathcal{C}(S,T,k,e)$ be the set containing all intervals $\mathit{lb}..\mathit{ub}\in\mathcal{B}[S,T,k]$ that are consistent with the endpoint discriminator:
$lb\leq k<ub$ when $e{=}0$, and $k{=}ub$ when $e{=}1$.
We define:

\vspace{-1em}
{\small
\begin{align}
\widehat{\mathcal{B}}[S,T,k,e] =
\begin{cases} \bigcap \mathcal{C}(S,T,k,e) & \text{if } \mathcal{C}(S,T,k,e) \neq \emptyset\\
\epsilon & \text{trap otherwise}
\end{cases}
\tag{\sc Bounds-Function}
\label{eq:bounds}
\end{align}
}

\noindent
Here, the operation fails if the result is the empty bounds, causing the strict operation to trap.
The intersection is contained in every admissible
interpretation and therefore cannot grant authority
outside any of them.
It may be smaller than the bounds of any individual
sub-object; the instantiation permits such conservative
restrictions of abstract authority.

This policy also resolves overlapping sub-objects.
For example, union members may admit different bounds for the same pointer type and offset.
Their common bounds are used, potentially rejecting
operations allowed by the abstract semantics.
The instantiation thus sacrifices some expressiveness rather than retaining sub-object history in each pointer.
That said, most pointers into arrays or \texttt{structs} are unambiguous, meaning that $|\mathcal{C}(S,T,k,e)| = 1$.
This means that the majority of \texttt{C} code will ordinarily be compatible, and exceptions may escape via \pnvi.

\myparagraph{Lock-step sub-object bounds}
In the instantiation, every strict capability will use the canonical bounds returned by $\widehat B$ for its representation and static type.
However, canonical bounds reconstruction is not itself sufficient to preserve strict capability authority: changing a pointer's static type or offset may select wider canonical bounds.
To address this, strict pointer-producing operations (other than allocation and recovery) explicitly enforce this invariant and monotonicity: the resulting canonical bounds must be contained within both the source capability's bounds and the bounds permitted by the corresponding abstract operation, or else the operation traps.
For example, casts and pointer arithmetic cannot restore previously lost authority, while member access must also respect the selected member's bounds.
These restrictions may reject operations allowed by the abstract semantics, but never grant additional authority.

The standard \texttt{memcpy}, \texttt{memmove}, and \texttt{memset} operations are treated specially: their checks use the original argument's authority rather than narrowing to the byte compatible sub-object.

\myparagraph{One-past-the-end pointers}
An allocation offset alone cannot always distinguish a pointer one-past a sub-object from a pointer to the beginning of an adjacent sub-object.
The discriminator $e$ preserves this distinction:
strict pointer arithmetic sets $e{=}1$ for a one-past
result and $e{=}0$ for an in-bounds result.
Reconstruction with $e{=}1$ considers only sub-objects ending at $k$, and a representation with no such interpretation traps.
This also allows backwards arithmetic from a one-past pointer to recover in-bounds canonical authority.
The discriminator distinguishes representations, not \texttt{C} pointer values.
Where \texttt{C} requires an endpoint pointer and a
pointer to the following sub-object to compare equal, strict pointer comparison preserves that equality despite their distinct representations.
Integer conversion instead exposes the representations, whose integer values need not compare equal.
The explicit discriminator also removes the need for the \emph{under-disambiguation} semantics of
\pnviaeudi.

\subsection{Computationally Unforgeable Representations}
\label{sec:frp-instantiation}

We showed how effective type, liveness, and sub-object bounds can be reconstructed without explicit per-pointer metadata.
However, this still leaves an important problem: how to recover the provenance $\pi$ of a pointer representation while preventing the fabrication of pointers to unrelated allocations?
We address this problem using \emph{computationally unforgeable} representations.

\myparagraph{Problem statement}
Consider a hypothetical implementation where $\widehat{\mathsf{repr}}(\pi,k,e)$ uses the underlying machine address $\mathit{addr}=\mathsf{base}(\pi){+}k$ (ignoring $e$ for now).
Such a representation is trivially \emph{forgeable}: given an address, integer arithmetic can readily construct addresses corresponding to adjacent objects in memory.
Consequently, a representation legitimately obtained from one allocation may be manipulated to construct a pointer into an unrelated allocation, even when the program has never obtained a pointer to the latter.
This is particularly problematic for \pnvi-style reconstruction, which permits arbitrary integer and byte-level manipulation of pointer representations.
We want to preserve such operations, including those that produce intermediate values that are not valid pointers, without allowing them to fabricate provenance for otherwise unavailable allocations.
Ordinary machine addresses do not provide this guarantee.

\myparagraph{Computationally unforgeable representations}
The fundamental challenge is that arbitrary integer representations
cannot provide absolute unforgeability.
Unlike hardware-tagged capabilities, ordinary integers cannot rely
on protected tags: intermediate integer values need not represent
a valid capability, and arbitrary bit patterns can always be
constructed (e.g., XOR linked-lists).
Nevertheless, while a valid representation can always be
\emph{guessed} in principle, we can make successful guessing
computationally infeasible by increasing the representation's
unpredictability.
We therefore introduce \emph{computational unforgeability}:
an adversary without access to a representation for an allocation
must have negligible probability of fabricating one for a
sufficiently large security margin.

Computational unforgeability can be realized using
\emph{Fully Randomized Pointers} (FRP)~\cite{phaye25frp}.
Despite the name, FRPs are best understood as
\emph{randomized handles} rather than machine addresses.
Each allocation is assigned an independently randomized
representation, unrelated to its underlying memory address or
to the representations of other allocations.
These handles retain pointer-like arithmetic: if $r$ represents
byte $k$ of an allocation, then $r{+}1$ represents byte $k{+}1$,
within the encoding's supported range.
Dereferencing a handle is not possible directly.
Crucially, fabricating a handle for an unavailable allocation
requires guessing a valid randomized
representation~\cite{phaye25frp}.

\myparagraph{Security margin}
The security of randomized representations can be parameterized
by entropy.
Suppose a representation uses $n$ bits of effective randomness,
and an attacker can make $q$ independent guesses with $V$
distinct targets.
The probability of successful forgery is therefore bounded by:

\vspace{-0.6em}
{\small
\[
    \Pr[\text{forgery}]
    \;\leq\;
    \min\left(1,\frac{qV}{2^n}\right)
\]
}

\noindent
For example, consider \emph{native} (64 bit) and
\emph{fat}\footnote{The fat width is justified by the 128 bit width commonly used by traditional capability systems.} (128 bit) pointer bit widths respectively.
Assuming that FRP can randomize all but 5 bits, then the effective security can be quantified as follows:

\vspace{-0.5em}
{\small
\[
\begin{array}{c|c|l}
\text{Pointer-width }(w) & \text{FRP-entropy }(n)
    & \mathrm{E}[\text{forgery}] \text{ in {\sc PacMan}-years} \\ \hline
64  & 59  & \qquad{\approx}24.8~\text{million years} \\
128 & 123 & \qquad{\approx}4.38{\times}10^{26}~\text{years}
\end{array}
\]
}

\noindent
Here, $\mathrm{E}[\text{forgery}]$ is the expected time until a successful forgery, and a {\em {\sc PacMan}-year} is a unit of guesses defined by running the {\sc PacMan} speculative
execution attack for 1 year at ${\sim}2.6\mathit{ms}$ per attempt~\cite{ravi2022pacman}, or ${\approx}1.21{\times}10^{10}$ guesses.
Thus, native 64 bit representations provide meaningful protection against forgery, while 128 bit representations can make forgery computationally infeasible under realistic threat models.\footnote{
Here, {\sc PacMan} already represents a very strong attacker model with unlimited attempts in a brute force-style attack.
Most realistic attackers will cause the program to trap (terminate) after a single failed guess.}

\myparagraph{Provenance recovery}
The high entropy of FRP means that provenance can be readily recovered from the representation, since each representation spans a specific randomized (non-overlapping) value range.
The underlying FRP infrastructure ensures that representations never collide.

\myparagraph{Strict capability recovery}
A valid FRP representation identifies both its allocation and its offset:
\[
  \widehat{\mathsf{repr}}(\pi,k,e)
    \longmapsto (\pi,k,e).
\]
Unlike explicit capability tags, this does not require protected metadata to accompany representations through intermediate computations or memory stores.
Exposed representations may undergo arbitrary integer and byte-level manipulation.

When a valid representation is recovered, the allocation identity, offset, and program's static type are sufficient to reconstruct and validate a full strict capability, as described in \autoref{sec:reconstructible}:
\[
\underbrace{\widehat{\mathsf{repr}}(\pi,k,e)}_{\pi,k,e}
\;+\;
\underbrace{\mathcal{T}[\pi],\Sigma[\pi]}_{T,\text{liveness}}
\;+\;
\underbrace{\widehat{\mathcal{B}}[S,T,k,e]}_{\mathit{lb}..\mathit{ub}}
\;+\;
\underbrace{S}_{\text{static type}}
\quad\Longrightarrow\quad
\langle\pi,S,T,\mathit{lb}..\mathit{ub},
        \mathsf{base}(\pi){+}k\rangle.
\]
Thus, no per-pointer provenance shadow metadata is required: the remaining strict capability metadata is recovered from the representation and program context.
Crucially, recovering a valid representation does not bypass strict capability validation.

\myparagraph{Example}
Consider two independently allocated objects, $\pi_p$ and
$\pi_q$, with FRP representations:
\[
\widehat{\mathsf{repr}}(\pi_p,0,0)
  = \texttt{\textcolor{green!50!black}{0xac8415a209566010}}
\qquad
\widehat{\mathsf{repr}}(\pi_q,0,0)
  = \texttt{\textcolor{red!50!black}{0x????????????????}}
\]
These values are adapted from the FRP example
in~\cite{phaye25frp} (Example 1).
Possession of the first representation permits integer
arithmetic to construct representations for offsets within
$\pi_p$.
For example, adding $4$ yields a representation that resolves
to $(\pi_p,4,0)$.
Recovery then checks liveness ($\Sigma[\pi_p]$), retrieves
effective type ($T = \mathcal{T}[\pi_p]$) and canonical
sub-object bounds ($\widehat{\mathcal{B}}[S,T,4,0]$),
before constructing a strict capability.
An incompatible type or inadmissible sub-object causes
recovery to fail (even though FRP resolution succeeds).
Unlike ordinary addresses, possession of the first
representation does not enable construction of a valid
representation for $\pi_q$: rather, the second independently
randomized value cannot feasibly be derived from the first.

\myparagraph{Guarantees and limitations}
Computational unforgeability is not equivalent to the
absolute protection provided by hardware capability tags.
An attacker who successfully guesses a valid handle may
obtain the corresponding provenance.
However, the security margin can be increased arbitrarily
in principle by using ever wider representations.
Conversely, representations already available to a computation
may be manipulated as ordinary integers and recovered,
subject to strict capability validation.
\tool therefore protects against fabrication of unavailable
provenance, rather than restricting the use of representations
already available to the program.

\subsection{Prototype Implementation}
\label{sec:implementation}

We prototype \tool to evaluate whether the strict capability programming model can be realized for existing \texttt{C} programs.
The prototype is intended as an executable model of the semantics, rather than a proposal for a production implementation or capability architecture.
In particular, our evaluation focuses on compatibility and the use of the strict/exposed boundary, rather than performance.
The prototype builds on \textsc{BlueFat}~\cite{phaye25frp} software implementation of FRP (configured for a 64 bit pointer width and one bit reserved for $e$), and adapts components of EffectiveSan~\cite{duck18effective} for dynamic type and sub-object reconstruction.
Conceptually, \textsc{BlueFat} also replaces the low fat pointer~\cite{duck16heap} mechanism used by EffectiveSan: both provide a compact pointer from which the containing allocation and offset can be identified, while FRP additionally provides the probabilistic provenance and exposure properties described in \autoref{sec:frp-instantiation}.
For each allocation $\pi$, we augment the \textsc{BlueFat} metadata with effective type $\mathcal{T}[\pi]$ information.
We also adapt EffectiveSan's layout machinery to implement $\widehat{\mathcal{B}}[S,T,k,e]$ as described in \autoref{sec:reconstructible}.

The remaining requirement is to enforce the strict operations from \autoref{fig:strict-semantics}.
These operations require the pointer's static type $S$, which is source information that is not recoverable from the FRP representation.
EffectiveSan already provides Clang/LLVM~\cite{llvm} compiler infrastructure for propagating this information to typed pointer arithmetic and memory operations, which we
adapt for \tool.
Thus, pointer arithmetic, casts, loads, and stores are interpreted relative to $S$ and enforce the corresponding strict rules, and operations in the exposed domain instead manipulate the FRP representation as ordinary data until recovery.
The resulting implementation follows directly from the reconstruction developed above:
\[
\underbrace{\widehat{\mathsf{repr}}(\pi,k,e)}_{\text{{\sc BlueFat}-style}}
\;+\;
\underbrace{\mathcal{T}[\pi],\Sigma[\pi],
             \widehat{\mathcal{B}}[S,T,k,e]}_{\text{EffectiveSan-style}}
\;+\;
\underbrace{S}_{\text{program}}
\quad\Longrightarrow\quad
\text{strict capability semantics}.
\]
No abstract capability tuple is materialized.
Instead, the prototype reconstructs and checks the required authority when executing strict operations.
This provides an executable implementation of the \tool abstraction with which we can investigate its central compatibility question:
whether \texttt{C} programs can operate predominantly in the strict domain, using explicit exposure and recovery only for idioms that genuinely require non-monotonic manipulation of pointer representations.

\myparagraph{Stack objects}
The prototype also supports stack-allocated objects.
Stack storage is allocated using ordinary LLVM \texttt{alloca} instructions, while the object's lifetime is instrumented to create and revoke the corresponding capability representation.
Thus, stack objects receive the same provenance, type, bounds, and liveness checks as heap objects, without requiring special stack handling.

\myparagraph{Limitations}
\tool enforces provenance, liveness, type, and sub-object bounds, but does not provide general information-flow security or protection against all forms of undefined behavior.
In particular, a program may legitimately possess authority to access data that it should not disclose, while vulnerabilities outside \tool's enforced invariants may disclose otherwise unavailable pointer representations.
However, this limitation is not unique to \tool: capability systems generally require complementary protections against unauthorized information flow and other classes of undefined
behavior.
Our prototype mitigates one important source of such disclosure by zero-initializing newly allocated memory.

\section{Evaluation}

We evaluate \tool against the following aspects:
\begin{enumerate}[leftmargin=*]
\item \label{case:idioms} compatibility and security of exposed representations against low-level \texttt{C} idioms;
\item \label{case:spec} whole source-level compatibility of strict capabilities with the SPEC benchmarks~\cite{henning06spec}; and
\item \label{case:study} case study porting requirements for full source-level compliance.
\end{enumerate}

\myparagraph{Goals}
The goal of this evaluation is to assess the \tool abstraction and programming model.
In particular, we investigate whether separating strict capabilities from exposed representations achieves the intended balance between strong language-level safety guarantees and compatibility with low-level \texttt{C} programming idioms.
Our experiments therefore focus on three questions (\ref{case:idioms})-(\ref{case:study})
which characterize the semantic and compatibility implications of \tool.

\myparagraph{Non-Goals}
This evaluation does not attempt to quantify the runtime overhead of the \tool prototype or establish production readiness.
Our prototype implements the additional checks required for effective types, sub-object bounds, and liveness in software, without optimizing their performance.
Efficient implementations with hardware support are considered in \autoref{sec:related}.

\myparagraph{Experimental Setup}
For testing low-level \texttt{C} idioms (\ref{case:idioms}), we curate a suite of microbenchmarks of standalone programs that test each idiom.
We compare \tool against the CHERI~\cite{cheri} purecap ABI as a representative of a traditional capability system.
The purecap experiments are run a CheriBSD/RISC-V instance under QEMU executing binaries compiled with the CHERI LLVM toolchain.
For (\ref{case:spec}) and (\ref{case:study}), we use the SPEC 2006 CPU \texttt{C} benchmarks compliant with the EffectiveSan runtime on which \tool is based.
All experiments are performed on a host system with an Intel Xeon Gold 6132 CPU (clocked at 2.60GHz) with 188GB of RAM.

{
\lstdefinestyle{c_idiom}{
  language=C,
  aboveskip=0pt,
  belowskip=-6pt,
  basicstyle=\ttfamily\fontsize{6.5pt}{7pt}\selectfont,
  keywordstyle=\bfseries\color{blue},
    keywordstyle=[1]\color{blue},
    morekeywords=[1]{
        void, char, int, long, unsigned, const, struct, 
        union, typedef, size_t, uint64_t, uintptr_t, uint8_t
    },
    keywordstyle=[2]\color{blue},
    morekeywords=[2]{
        if, while, return, break, sizeof, assert
    },
    keywordstyle=[3]\color{purple},
    morekeywords=[3]{
        va_list, va_start, va_arg, allocate, call, 
        has_zero, __attribute__
    },
    commentstyle=\color{green},
    stringstyle=\color{brown},
    numberstyle=\color{red},
    identifierstyle=\color{black},
    literate=
        {~}{{\textcolor{black}{\textasciitilde}}}{1}
        {&}{{\textcolor{black}{\&}}}{1}
        {^}{{\textcolor{black}{\^}}}{1}
        {|}{{\textcolor{black}{|}}}{1}
        {=}{{\textcolor{black}{=}}}{1}
        {+}{{\textcolor{black}{+}}}{1}
        {-}{{\textcolor{black}{-}}}{1}
        {*}{{\textcolor{black}{*}}}{1}
        {/}{{\textcolor{black}{/}}}{1},
  commentstyle=\itshape\color{gray},
  stringstyle=\color{red},
  breaklines=true,
  breakatwhitespace=true,
  framexleftmargin=-1pt,
  framexrightmargin=-218pt,
  framextopmargin=-5pt,
  framexbottommargin=-2pt,
  backgroundcolor=\color{gray!8},
  tabsize=4,
  showspaces=false,
  showstringspaces=false,
  showtabs=false,
  rulecolor=\color{black},
  captionpos=b,
  numbers=none,
  columns=flexible,
  escapechar=@,
  upquote=true,
  morecomment=[l]{//},
  sensitive=true,
  morekeywords={typeof, _Bool, _Complex, _Imaginary, restrict, inline, _Noreturn, _Static_assert},
}

\lstset{style=c_idiom}
\lstset{ literate={~}{{\raisebox{0.5ex}{\texttildelow}}}{1} }

\begin{table}[]
    \footnotesize
    \caption{Representative provenance exposure cases under \pnvi (E.1-E.4), known idioms that rely on pointer exposure (I.1-I.6), failing Juliet test case examples (J.1, J.2), and a Juliet test summary (T.1-T.4).
    }
    \label{tab:c-idioms}
    \centering
    \begin{tabular}{|l|p{1.4cm}|p{6cm}|p{4.5cm}|}
    \hline
    \rowcolor{red!10}
      & {\em Idiom}
      & {\em Representative \texttt{C} Code Snippet}
      & {\em Description} \\ \hline

    E.1 & Pointer-integer round-trips &
\vspace{-0.7em}
\begin{lstlisting}
int *p = &x;
uintptr_t r = (uintptr_t)p;
int *q = (int *)r;
\end{lstlisting}
    & The pointer-to-integer cast exposes \texttt{x}.
      The reverse cast recovers \texttt{x}'s provenance from the integer representation. \\
    \hline

    E.2 & Reading raw pointer bytes &
\vspace{-0.7em}
\begin{lstlisting}
int *p = &x, *q;
size_t n = sizeof p;
uint8_t *s = (void *)&p;
uint8_t *d = (void *)&q;
while (n--) *d++ = *s++;
\end{lstlisting}
    & Reading \texttt{p}'s representation as bytes
      exposes \texttt{x}. The copied representation
      allows \texttt{q} to recover \texttt{x}'s provenance. \\
    \hline

    E.3 & Union type punning &
\vspace{-0.7em}
\begin{lstlisting}
union { int *p; uintptr_t u; } v;
v.p = &x;
int *q = (int *)v.u;
\end{lstlisting}
    & Reading a pointer representation through
      integer member \texttt{v.u} exposes \texttt{x}.
      The resulting \texttt{q} recovers \texttt{x}'s provenance. \\
    \hline
    E.4 & Pointer I/O &
\vspace{-0.7em}
\begin{lstlisting}
int *p = &x, *q;
fprintf(stream, "%p", p);
rewind(stream);
fscanf(stream, "%p", &q);
\end{lstlisting}
    & Printing/writing \texttt{p}'s representation with I/O (e.g., \texttt{printf}'s \texttt{\%p}) will expose \texttt{x}.
      Reading back the representation recovers \texttt{x}'s provenance in \texttt{q}. \\
    \hline

\rowcolor{blue!10}
I.1 & Container-of &
\vspace{-0.7em}
\begin{lstlisting}
struct S { int x, y; } s;
uintptr_t r = (uintptr_t)&s.y;
size_t k = offsetof(struct S, y);
struct S *q = (struct S *)(r - k);
\end{lstlisting}
& Uses E.1 to recover enclosing-object provenance and
  bounds from a member pointer. \\
\hline
\rowcolor{blue!10}
I.2 & XOR linked-lists &
\vspace{-0.7em}
\begin{lstlisting}
struct N a, b, c;
b.link = (uintptr_t)&a ^ (uintptr_t)&c;
struct N *q = (struct N *)(b.link ^ (uintptr_t)&a);
\end{lstlisting}
& Uses E.1 to encode two pointers as their XOR and
  recover one using the other. \\
\hline
\rowcolor{blue!10}
I.3 & Tagged pointers &
\vspace{-0.7em}
\begin{lstlisting}
uintptr_t r = (uintptr_t)&x;
uintptr_t tagged = r | 1u;
int *q = (int *)(tagged & ~(uintptr_t)1);
\end{lstlisting}
& Uses E.1 to store metadata in unused pointer bits
  and recover the original pointer. \\
\hline
\rowcolor{blue!10}
I.4 & Pointer obfuscation &
\vspace{-0.7em}
\begin{lstlisting}
uintptr_t key = 0x1234;
uintptr_t r = (uintptr_t)&x ^ key;
int *q = (int *)(r ^ key);
\end{lstlisting}
& Uses E.1 to reversibly encode a pointer as an
  integer before recovery. \\
\hline
\rowcolor{blue!10}
I.5 & Pointer encryption / decryption &
\vspace{-0.7em}
\begin{lstlisting}
int *p = &x, *q;
unsigned char buf[sizeof p];
for (size_t i = 0; i < sizeof p; i++)
    buf[i] = ((uint8_t *)&p)[i] ^ 0x5a;
for (size_t i = 0; i < sizeof q; i++)
    ((uint8_t *)&q)[i] = buf[i] ^ 0x5a;
\end{lstlisting}
& Uses E.2 to obfuscate a pointer
  representation at the byte-level. \\
\hline
\rowcolor{blue!10}
I.6 & Self-pipe pointer I/O &
\vspace{-0.7em}
\begin{lstlisting}
int *p = &x, *q, fds[2];
pipe(fds);
write(fds[1], &p, sizeof p);
read( fds[0], &q, sizeof q);
\end{lstlisting}
& Uses E.4 to serialize a pointer through a
  self-pipe. \\
\hline

\rowcolor{orange!12}
J.1 & Expired stack object (CWE843) &
\vspace{-0.7em}
\begin{lstlisting}
void *data;
{
    int x = 8;
    data = &x;
}
printIntLine(*(int *)data);
\end{lstlisting}
& \textbf{Rejected:} \texttt{x}'s lifetime has ended.
  Fix by extending its scope to include the access. \\
\hline

\rowcolor{orange!12}
J.2 & \texttt{void} pointer round-trip (CWE588) &
\vspace{-0.7em}
\begin{lstlisting}
struct S { int x, y; } s;
void *p = &s;
struct S *q = (struct S *)p;
\end{lstlisting}
& \textbf{Rejected:} The instantiation narrows
  \texttt{p}'s authority to the \texttt{s.x} sub-object, rather than \texttt{s}.
  Use \texttt{uintptr\_t} (\pnvi escape hatch) and checked recovery instead. \\
\hline

\rowcolor{green!12}
T.1 & CWE121/122 & Object bounds error & 144/144 good pass, 144/144 bad detected \\
\hline
\rowcolor{green!12}
T.2 & CWE188        & Sub-object bounds error & 18/18 good pass, 18/18 bad detected \\
\hline
\rowcolor{green!12}
T.3 & CWE416 & Use-after-free
    & 118/118 good pass, 82/118 bad detected \\
\hline
\rowcolor{green!12}
T.4 & CWE588/843 & (Ported) Type error & 102/102 good pass, 102/102 bad detected \\
\hline
    \end{tabular}
\end{table}
}

\subsection{Low-level C Idiom Compatibility}

Our goal is to demonstrate that \tool supports common low-level \texttt{C} idioms that are permitted under a modern \pnvi provenance semantics.
Our microbenchmark suite evaluates 4 representative {\em exposure idioms} (E.1-E.4) and 6 {\em low-level \texttt{C} idioms} (I.1-I.6) listed in Table~\ref{tab:c-idioms}, covering integer-pointer casts, raw memory operations, and pointer I/O.
Each idiom is tested as a standalone program compiled and executed under purecap and \tool. 

The results indicate that \tool correctly implements low-level \texttt{C}/\pnvi behavior on all 10/10 benchmarks.
In contrast, purecap fails outright for $\{\text{E.2}, \text{E.4}, \text{I.2}, \text{I.3}, \text{I.4}, \text{I.5}, \text{I.6} \}$ due to the inability to track provenance/authority through pointer byte manipulation, I/O, and arbitrary integer operations.
Purecap conditionally supports some basic pointer-integer round trips $\{\text{E.1}, \text{E.3}\}$ by treating (\texttt{uintptr\_t}) as a special provenance-carrying integer type~\cite{zaliva24cheri}.\footnote{In CHERI \texttt{C}, the (\texttt{u})\texttt{intptr\_t} types are essentially 128-bit capabilities rather than ordinary integer types.
As such, the same limitations apply, and combinations of certain operations like $(p\texttt{+}100001)\texttt{-}100000$ may still clear the capability tag~\cite{zaliva24cheri}.}
However, these benchmarks also fail if (\texttt{uintptr\_t}) were substituted with an ordinary integer type, such as (\texttt{uint64\_t}), whereas \tool is unaffected.
Likewise, purecap could pass E.2 if the program used \texttt{memcpy} instead as a special-case, although general pointer byte manipulation would still be rejected.
Finally, purecap could also pass I.1 if sub-object bounds narrowing were disabled, see \autoref{sec:example}.

These results demonstrate that representation-level compatibility need not require weakening ordinary strict capability operations.
All ten idioms execute under \tool, while exposure remains non-dereferenceable and recovery remains checked.

\myparagraph{Strictness validation}
We evaluate \tool on selected \texttt{C} programs from the Juliet Test Suite~\cite{juliet}, covering whole object bounds errors (T.1, CWE121/122), sub-object bounds errors (T.2, CWE188), use-after-free errors (T.3, CWE416), and type confusion/errors (T.4, CWE588/843).
Each program is compiled with \texttt{-O2 -g -fno-inline} and executed separately along both good path (i.e., ostensibly fault-free) and bad paths (i.e., contains the fault).
The results are shown in \autoref{tab:c-idioms}.
Interestingly, not all Juliet tests initially pass, and common failures include: compiler optimization (\texttt{-O2}) removing the error before the \tool instrumentation pass, \tool enforcing monotonicity restrictions stronger than the \texttt{C} standard, and unrelated bugs in specific Juliet tests.
Initially, 68/68 of the CWE843 good tests failed due to \tool detecting an incidental lifetime bug (illustrated by \autoref{tab:c-idioms} J.1), and 13/34 CWE588 good tests violate the instantiation's treatment of (\texttt{void *}) bounds.
Specifically, the (\ref{eq:bounds}) conservatively selects the narrowest sub-object bounds by design, which can be narrower than intended (illustrated by \autoref{tab:c-idioms} J.2).
The former is undefined \texttt{C} behavior, while the latter is incompatible with the instantiation's monotonicity enforcement.
Fixing the CWE843 lifetime bugs and porting CWE588 (by substituting (\texttt{void *}) with \texttt{uintptr\_t}) resolves these issues, with 86/86 revised good tests passing.
Likewise, \tool correctly traps all bad tests except for 36 missed detections in T.3 due to compiler optimization.

\subsection{SPEC Compatibility}
\label{sec:spec-compatibility}

We evaluate \tool against the 12 SPEC CPU2006 \texttt{C}
benchmarks~\cite{henning06spec}.
Although dated, this is the same version of SPEC used in the EffectiveSan evaluation~\cite{duck18effective}, on which our prototype is based.
Our baseline uses the EffectiveSan-patched SPEC sources, which remove several custom memory allocators (CMAs).\footnote{
\url{https://github.com/GJDuck/EffectiveSan/blob/30a34f5018280ce7273db5490c2a06da65adfe75/spec2006/cpu2006.patch}}
We investigate whether these programs can be adapted to satisfy \tool's strict capability invariants.
We instrument the baseline to identify violations of pointer arithmetic, member access, casts, dereferences, and integer-to-pointer conversions.
We use Codex~\cite{codex} to generate source-level repairs to eliminate violations while preserving program behavior.

\myparagraph{Baseline compatibility}
\autoref{tab:spec-sw} reports violations observed with the SPEC \texttt{test} inputs.
Violations occur at 17,583 of 38,368 checked lines (45.83\%) overall, and most significantly with 72.87\% in \texttt{403.gcc}, 31.90\% in \texttt{400.perlbench}, and 36.02\% in \texttt{433.milc}.
These figures measure the breadth of observed violations, not the number of independent defects or required repairs.
For example, a single incompatible allocation pattern in \texttt{403.gcc} affects thousands of consumers.

\myparagraph{Porting results}
All 12 benchmarks were adapted to pass their official \texttt{test} workloads under \tool's instrumentation.
Across 26 input executions, the ported programs produced the expected outputs with zero failures among 105.2 billion executed encoded checks.

The \emph{Modified} columns of \autoref{tab:spec-sw}
report the cumulative source patches: 146 files,
with 3,048 lines added and 1,814 removed.
The largest ports were \texttt{400.perlbench} (36 files, 2,452 changed lines) and \texttt{403.gcc} (58 files, 1,371 changed lines).
Only two benchmarks required no source changes.
Widespread violations need not correspond to
equally widespread independent repairs, although some ports require substantial restructuring.
These results establish compatibility for the tested execution paths, not necessarily the complete compatibility of the programs for all inputs.
The reported patches are the cumulative result of AI-assisted porting and are not necessarily minimal or optimal repairs.
Nevertheless, the exercise demonstrates how porting to strict capabilities is feasible even for highly non-compliant code.

\myparagraph{Comparison with the purecap ABI} Purecap supports most SPEC CPU2006 \texttt{C} benchmarks with recompilation alone, although \texttt{400.perlbench}, \texttt{403.gcc}, and \texttt{429.mcf} were reported as purecap incompatible~\cite{richardson2020cheri}.
All three are included in our successful ports.
Purecap generally imposes fewer restrictions on pointer arithmetic, types, and sub-object bounds than \tool.
Conversely, \tool's \pnvi escape hatch accommodates certain representation-manipulating idioms through checked recovery.
Our results show that existing code can nevertheless be ported to strict capabilities.

\begin{table*}[t]
\centering
\caption{
Baseline strict-capability violations in the SPEC CPU2006 \texttt{C} benchmarks using \texttt{test} inputs.
\emph{Locs} counts distinct instrumented locations (source lines) executed during the workload.
Each percentage denotes the fraction of these lines at which at least one violation of the indicated operation kind occurred.
Operation kinds may overlap, and {\em Failed} counts the union of violating locations.
\emph{Status} indicates successful execution of the official SPEC test workloads, with matching expected outputs and no observed encoded-check failures after porting.
\emph{Modified} reports the number of modified source files and lines removed/added relative to the EffectiveSan-patched baseline.
Here, $-/+$ means lines removed/added.
}
\label{tab:spec-sw}
\footnotesize
\setlength{\tabcolsep}{4pt}
\begin{tabular}{lrrrrrrrrrc}
\toprule
\multirow{2}[1]{*}{Benchmark}
  & \multirow{2}[1]{*}{Locs}
  & \multicolumn{5}{c}{Violation by Operations (\% of Total)}
  & \multirow{2}[1]{*}{Failed}
  & \multicolumn{2}{c}{Modified}
  & \multirow{2}[1]{*}{Status} \\
\cmidrule(lr){3-7}
\cmidrule(lr){9-10}
  & & \autoref{rule:arith} & \autoref{rule:member}
  & \autoref{rule:cast} & \autoref{rule:deref}
  & {\sc Int2Ptr} & & Files & Lines & \\
\midrule
400.perlbench
  & 10,470 & 1.05 & 25.03 & 15.21 & 8.76 & $<0.01$
  & \cellcolor{red!31} 31.90\% & 36 & $\textcolor{red!50!black}{-1009}/\textcolor{green!40!black}{+1443}$ & \textcolor{green!40!black}{Pass} \\
401.bzip2
  & 812 & 6.53 & 0.00 & 1.48 & 6.53 & 0.00
  & \cellcolor{red!19} 8.00\% & 3 & $\textcolor{red!50!black}{-24}/\textcolor{green!40!black}{+35}$ & \textcolor{green!40!black}{Pass} \\
403.gcc
  & 18,922 & 0.47 & 35.64 & 26.68 & 68.25 & 0.02
  & \cellcolor{red!51} 72.87\% & 58 & $\textcolor{red!50!black}{-374}/\textcolor{green!40!black}{+997}$ & \textcolor{green!40!black}{Pass} \\
429.mcf
  & 220 & 0.45 & 0.00 & 3.64 & 0.00 & 0.00
  & \cellcolor{red!17} 4.09\% & 3 & $\textcolor{red!50!black}{-30}/\textcolor{green!40!black}{+34}$ & \textcolor{green!40!black}{Pass} \\
433.milc
  & 669 & 0.45 & 5.53 & 11.21 & 24.51 & 0.00
  & \cellcolor{red!33} 36.02\% & 13 & $\textcolor{red!50!black}{-195}/\textcolor{green!40!black}{+211}$ & \textcolor{green!40!black}{Pass} \\
445.gobmk
  & 2,440 & 0.12 & 0.08 & 0.08 & 0.04 & 0.00
  & \cellcolor{red!15} 0.29\% & 4 & $\textcolor{red!50!black}{-2}/\textcolor{green!40!black}{+21}$ & \textcolor{green!40!black}{Pass} \\
456.hmmer
  & 498 & 0.00 & 0.00 & 0.00 & 0.00 & 0.00
  & \cellcolor{green!15} 0.00\% & 0 & $-$ & \textcolor{green!40!black}{Pass} \\
458.sjeng
  & 363 & 0.00 & 0.00 & 0.00 & 0.00 & 0.00
  & \cellcolor{green!15} 0.00\% & 0 & $-$ & \textcolor{green!40!black}{Pass} \\
462.libquantum
  & 123 & 0.81 & 0.00 & 1.63 & 0.00 & 0.00
  & \cellcolor{red!16} 2.44\% & 1 & $\textcolor{red!50!black}{-4}/\textcolor{green!40!black}{+2}$ & \textcolor{green!40!black}{Pass} \\
464.h264ref
  & 2,022 & 0.00 & 0.05 & 0.00 & 0.05 & 0.00
  & \cellcolor{red!15} 0.05\% & 1 & $\textcolor{red!50!black}{-1}/\textcolor{green!40!black}{+2}$ & \textcolor{green!40!black}{Pass} \\
470.lbm
  & 188 & 0.00 & 5.32 & 2.66 & 3.19 & 0.00
  & \cellcolor{red!19} 8.51\% & 2 & $\textcolor{red!50!black}{-28}/\textcolor{green!40!black}{+41}$ & \textcolor{green!40!black}{Pass} \\
482.sphinx3
  & 1,641 & 4.88 & 0.00 & 1.65 & 3.47 & 0.00
  & \cellcolor{red!18} 6.89\% & 25 & $\textcolor{red!50!black}{-147}/\textcolor{green!40!black}{+262}$ & \textcolor{green!40!black}{Pass} \\
\midrule
\textbf{Total}
  & 38,368 & 0.88 & 24.54 & 17.65 & 36.78 & 0.01
  & \cellcolor{red!38} 45.83\% & 146 & $\textcolor{red!50!black}{-1814}/\textcolor{green!40!black}{+3048}$ & \textcolor{green!40!black}{12/12} \\
\bottomrule
\end{tabular}%
\end{table*}

\subsection{SPEC Porting Case Studies}
\label{sec:spec-cases}

The SPEC ports illustrate two distinct approaches to resolving strict-capability violations: {\em repairing object representations} to satisfy strict invariants, and using the {\em \pnvi escape hatch} when valid authority must be recovered after representation manipulation.
We consider two case studies.

\myparagraph{Repairing object representations}
\texttt{403.gcc}'s garbage collector originally carved compact tree and RTL objects from byte-typed allocation pages.
Under \tool, such casting does not establish the effective type of a \texttt{tree\_node} or
\texttt{rtx\_def}, resulting in thousands of type violations at consumer locations.
The port introduces genuinely typed allocations and integrates these with \texttt{gcc}'s existing garbage collection machinery.
The port also replaces certain variable-length and overlapping representations with typed operands and separately allocated vectors.
This preserves the relevant object types and lifetimes rather than relying on casts to manufacture authority.

The \texttt{gcc} benchmark also deliberately creates out-of-bounds pointers to allow absolute indexing~\cite{duck16heap}, which strict capabilities prohibit.
The port instead allocates the unused prefix:

{\small
\begin{Verbatim}[commandchars=\\\{\}]
        \textcolor{darkred}{-} qty_table = xmalloc((max_qty - max_reg) * \textcolor{darkyellow}{sizeof}(*qty_table));
        \textcolor{darkred}{-} qty_table -= max_reg;
        \textcolor{darkgreen}{+} qty_table = xmalloc(max_qty * \textcolor{darkyellow}{sizeof}(*qty_table));
\end{Verbatim}
}

\myparagraph{Checked authority recovery}
\texttt{433.milc} passes typed matrix and vector objects through generic (\texttt{char *}) interfaces.
Such conversions can narrow the reconstructed bounds, so casting the resulting pointer back to the original type can violate \tool's strict containment rule.
The port instead performs the intervening address calculation as integer arithmetic and uses integer-to-pointer recovery at the typed consumer.
Recovery reconstructs authority from the live allocation and still checks its effective type and bounds.
Unlike a strict cast, this need not preserve the narrowed source bounds.
This illustrates the intended role of the \pnvi escape hatch: recovering valid authority, not creating a typed object in untyped storage.

\myparagraph{Discussion}
Together, these cases show how the escape hatch complements, rather than replaces, strict typing: some idioms require genuine representation changes, while others can retain their representation-level behavior through checked recovery.
Overall, all 12 \texttt{C} SPEC benchmarks were were successfully adapted without abandoning strict capability enforcement or removing the low-level idioms wholesale.
Where non-monotonic operations are required, the changes are local and confined, and use the existing \texttt{C} and \pnvi semantics.
This validates the dual capability abstraction and design.

\section{Related Work}\label{sec:related}

\myparagraph{Memory-safety enforcement}
A wide range of software and hardware techniques enforce memory-safety
properties in \texttt{C}/\texttt{C++}.
AddressSanitizer~\cite{asan}, Oscar~\cite{dang2017oscar},
and hardware-assisted schemes such as
HeapCheck~\cite{saileshwar2022heapcheck} target spatial or temporal
memory errors.
Fat-pointer systems~\cite{safe-c,necula02ccured,cyclone},
low-fat pointers~\cite{duck16heap,duck17stack},
SoftBound~\cite{nagarakatte09softbound}, and
Intel MPX~\cite{intel23manual} associate pointers with bounds,
either explicitly or through auxiliary metadata.
These approaches provide different safety and compatibility
trade-offs, but do not generally enforce the combination of
provenance, type correctness, and sub-object bounds targeted by
\tool.

EffectiveSan~\cite{duck18effective} is particularly relevant: it uses allocation-time type metadata and a type-layout function to check effective types and sub-object bounds.
Given an allocation's dynamic type, a pointer's static type, and an interior offset, it reconstructs admissible sub-object bounds.
However, the two systems resolve ambiguity differently: EffectiveSan favors {\em permissive bounds}, effectively taking the \emph{union} of admissible interpretations to reduce false positives, whereas \tool takes their \emph{intersection} to prevent authority widening, as illustrated as follows:
\[
\widehat{B}_{\tool} = \bigcap_{B\in\mathcal{C}} B
\qquad\qquad \text{versus} \qquad\qquad
\widehat{B}_{\mathrm{EffectiveSan}} \approx \bigcup_{B\in\mathcal{C}} B
\]
This reflects the different goals: EffectiveSan is a bug-detection sanitizer which aims to minimize false positives at the cost of some strictness, whereas \tool must conservatively enforce strict capability authority---even if at the cost of rejecting otherwise valid operations.

\myparagraph{Hardware capabilities and CHERI \texttt{C}}
Hardware capability systems, including CHERI~\cite{cheri}
and Capstone~\cite{yu2023capstone}, protect pointer authority
through capability metadata and restricted derivation operations.
CHERI capabilities encode bounds and permissions, with hardware
tags preventing arbitrary byte sequences from becoming valid
capabilities.
This provides strong spatial protection, but creates tension
with \texttt{C} idioms that manipulate pointer representations
as ordinary integers or bytes.
Moreover, capability bounds do not inherently enforce
effective types or sub-object bounds.

Zaliva et al.~\cite{zaliva24cheri} investigate these tensions in detail, developing an executable and mechanized semantics of CHERI \texttt{C} based on Cerberus~\cite{memarian2023cerberus} and \pnviaeudi.
Their work shows how CHERI-specific restrictions can be integrated into a provenance-aware \texttt{C} semantics.
In particular, CHERI \texttt{C} represents \texttt{uintptr\_t} as a capability-carrying integer type, preserving authority across supported pointer-integer operations rather than treating it as an ordinary integer.
Conversely, byte-wise modification of a capability representation can invalidate its use for memory access, even when the resulting bytes appear unchanged.
Their semantics also reflects CHERI \texttt{C}'s default policy of not enforcing sub-object bounds, in part to accommodate idioms such as \texttt{container\_of}.

\tool takes a different approach to these compatibility constraints.
Strict capabilities enforce provenance, type, and sub-object bounds, while exposed representations may undergo ordinary integer and byte-level manipulation without directly authorizing memory access.
Conversion back to a strict capability reconstructs and validates the required authority.
Thus, compatibility with representation-level idioms need not require weakening the invariants of ordinary strict pointer operations.
Unlike CHERI's protected capability representations, \tool does not rely on preserving embedded authority metadata through arbitrary representation transformations.

\myparagraph{Other pointer provenance proposals}
The \texttt{C++} proposal P2434
(\emph{Nondeterministic Pointer Provenance})%
~\cite{Herring2026P2434R5} explores a different
approach to provenance reconstruction.
Rather than explicitly recording which allocations have
been exposed, it uses nondeterminism in pointer
representations to constrain what programs may reliably
infer from integer values.
This shares with \tool the intuition that arbitrary
address guessing should not reliably manufacture usable pointers, but the mechanisms differ.
P2434 expresses this constraint through language semantics, whereas \tool uses randomized representations in its concrete realization.
Moreover, P2434 does not yet fully specify all cases of partial byte-wise modification of pointer objects, whereas such representation-level operations are a central motivation for \tool.

\myparagraph{Hardware implementation feasibility}
While this paper focuses on the \tool abstraction and programming model with a software prototype, existing work provides evidence that hardware acceleration is feasible.
{\sc GreenFat}~\cite{phaye25frp} demonstrates a hardware-oriented FRP implementation evaluated using architectural simulation, including accelerated resolution of randomized pointers.
In-Fat Pointer~\cite{xu2021infat} demonstrates hardware-assisted sub-object bounds checking, using a two-stage metadata lookup and bounds-narrowing mechanism inspired by EffectiveSan.
These systems provide precedents for two central components of \tool: provenance recovery from randomized representations and type-aware sub-object bounds enforcement.
A complete hardware implementation of \tool would require integrating these mechanisms with effective-type and liveness metadata, canonical bounds reconstruction, and the lock-step restrictions on strict capability operations.
We believe this integration will present substantial engineering challenges, and an optimized implementation remains to be designed and evaluated.
Nevertheless, the existing mechanisms suggest that hardware support is a plausible direction, rather than an inherent obstacle to the proposed abstraction.

\section{Conclusion}\label{sec:conclusion}

Capability systems have traditionally focused on enforcing
provenance, liveness, and allocation bounds. Yet these
guarantees alone are insufficient to prevent type confusion
and sub-object bounds violations. Strengthening capabilities
to enforce these additional invariants promises greater
security, but also intensifies a longstanding compatibility
problem: low-level \texttt{C} programs routinely manipulate
pointer representations in ways that strict capability
derivation cannot permit.

This paper presented \tool, a dual capability abstraction
that reconciles these requirements by separating
\emph{strict capabilities} from \emph{exposed representations}.
Ordinary pointer operations preserve strict authority,
including type and sub-object bounds, while explicit
exposure permits representation-level manipulation followed
by checked recovery under \pnvi. Crucially, this escape hatch
is not an arbitrary weakening of capability enforcement:
it follows the permissive recovery semantics developed to
accommodate low-level \texttt{C} idioms, without weakening
the invariants of ordinary strict capability operations.
The significance of this separation is that
\emph{strictness and compatibility need not be opposing goals}.
By confining permissive operations to an explicit boundary,
\tool allows programs to remain in the strict domain wherever
possible, resorting to exposure only when necessary.
Programs that avoid exposure altogether can operate entirely
under strict capability discipline.

Our evaluation demonstrates the viability of this approach: \tool supports low-level pointer idioms and the adaptation of existing \texttt{C} programs, while detecting type, lifetime, and sub-object violations beyond the guarantees of allocation-bounded capabilities.
More broadly, \tool suggests that stronger capability
enforcement need not require abandoning the flexibility of systems languages; instead, that flexibility can be isolated behind an explicit, validated recovery boundary.

\newpage

\bibliographystyle{ACM-Reference-Format}
\bibliography{main}

\end{document}